\documentclass[aps,prd,preprintnumbers,showpacs]{revtex4}
\usepackage[dvips]{graphicx}

\begin{document}

%
%

\eprint{Nisho-2-2021}
\title{Enhanced Radiation ? from ( Super ) Conductor by Dark Matter Axion}
\author{Aiichi Iwazaki}
\affiliation{International Economics and Politics, Nishogakusha University,\\ 
6-16 3-bantyo Chiyoda Tokyo 102-8336, Japan }   
\date{Nov. 4, 2021}
\begin{abstract}
In our previous paper we have shown
that enhanced radiations arise from ( super ) conductor by dark matter axion under strong magnetic field.
We have recalculated the radiation by carefully treating boundary conditions between vacuum and the conductor.
We show that the radiation is never enhanced. We interpret it such that 
electromagnetic field induced by axion collides the conductor and it is reflected by the conductor.
The reflecting wave is the radiation from the conductor. It is reasonable that
the amplitude of the reflection wave is identical to that of the incoming wave, i.e. radiation induced by axion.
\end{abstract}
\hspace*{0.3cm}

\hspace*{1cm}

\maketitle

We have shown in previous papers \cite{iwa} that radiations by axion from ( super ) conductor under strong magnetic field are enhanced,
compared with those in cavity experiments. Here we present a result obtained by recalculating
them, carefully treating boundary conditions between vacuum and conductor. The result shows that
radiations are never enhanced. We present a physical reason why they are not enhanced.
Out result also holds for superconductor, although we describe it in the case of normal conductor.

\vspace{0.1cm}  
The system of dark matter axion and electromagnetic fields in conductor under external spatially uniform and static magnetic field $\vec{B}$,
is described by the following modified Maxwell equations of motion,

\begin{eqnarray}
\label{metal}
&&\vec{\partial}\cdot\vec{E}(\mbox{inside})+g_{a\gamma\gamma}\vec{\partial}\cdot\Big(a(\vec{x},t)(\vec{B}(\mbox{inside})+\vec{B})\Big)=0, \nonumber \\ 
&&\vec{\partial}\times \Big(\vec{B}(\mbox{inside})-g_{a\gamma\gamma} a(\vec{x},t)\vec{E}(\mbox{inside})\Big)- 
\partial_t\Big(\vec{E}(\mbox{inside})+g_{a\gamma\gamma} a(\vec{x},t)\big(\vec{B}(\mbox{inside})+\vec{B}\big)\Big)=\vec{J},  \nonumber  \\
&&\vec{\partial}\cdot\vec{B}(\mbox{inside})=0, \quad \vec{\partial}\times \vec{E}(\mbox{inside})+\partial_t \vec{B}(\mbox{inside})=0.
\end{eqnarray}
with $\vec{J}=\sigma \vec{E}(\mbox{inside})$ ( $\sigma$ denotes electrical conductivity ),
along with the equation of motion of the axion,

\begin{equation}
(\partial_t^2-\vec{\partial}^2+m_a^2)a(t,\vec{x})=g_{a\gamma\gamma}\vec{E}(\mbox{inside})\cdot\big(\vec{B}(\mbox{inside})+B\big)
\end{equation} 
where we have used the equations $\vec{\partial}\times \vec{B}=0$ and $\vec{\partial}\cdot \vec{B}=\partial_t\vec{B}=0$.
We assume the trivial permeability $\mu=1$ and electric permittivity $\epsilon=1$.  
The equations of motion are derived by using Lagrangian 
$L_{aEB}=g_{a\gamma\gamma} a(\vec{x},t)\vec{E}(\mbox{inside})\cdot \big(\vec{B}(\mbox{inside})+\vec{B}\big)$
describing the coupling between axion $a(\vec{x},t)$, electric field $\vec{E}(\mbox{inside})$ and magnetic field $(\vec{B}(\mbox{inside})+\vec{B})$, 
and Lagrangian of electromagnetic fields as well as that of axions neglecting potential terms such as $(a/f_a)^4$ with 
the decay constant $f_a\sim 10^{12}$GeV.

The coupling constant $g_{a\gamma\gamma}\propto 1/f_a$ is extremely small such as $g_{a\gamma\gamma}\vec{E}\cdot\vec{B} \ll m_a^2a$.
Thus, the dark matter axion is almost free with small momentum $\vec{k} \ll m_a$. Thus, 
$a(\vec{x},t)$ can be described as $a(t)=a_0\cos(m_at)$.
Because the density $\rho_a\simeq m_a^2\overline{a(t)^2}=m_a^2a_0^2/2$ of the dark matter axion is 
extremely small such as $g_{a\gamma\gamma}a_0=g_{a\gamma\gamma}\sqrt{2\rho_a/m_a^2} \ll 1$, its coupling with electromagnetic
fields can be treated as a small perturbation. Thus, the fields $\vec{E}(\mbox{inside})$
and $\vec{B}(\mbox{inside})$ are of the order of
$g_{a\gamma\gamma}a(t)$.

When we apply the equations to fields in vacuum, we should note that there is no electric current $\vec{J}=\sigma \vec{E}(\mbox{inside})$
in vacuum.
We denote the vacuum solutions of electric and magnetic fields as $\vec{E}(\mbox{outside})$ and $\vec{B}(\mbox{outside})$.
As is well known, the boundary conditions on the electric and magnetic fields must be satisfied on the surface of the conductor.
They demand that the parallel components of electric and magnetic fields to the surface are continuous.  
The equations in eq(\ref{metal}) along with the boundary conditions are basic tools for the discussion given below.

\vspace{0.3cm} 

It is well known that
dark matter axion $a(t)=a_0\cos(m_at)$ generates oscillating electric field 
$\vec{E}_a=-g_{a\gamma\gamma}a(t)\vec{B}$  in vacuum under strong magnetic field $\vec{B}$.
The solution is easily obtained from the above equations with denoting $\vec{E}(\mbox{outside})=\vec{E}_a$.
Although such an electric field is generated also in normal conductor, the oscillating electric field 
$\vec{E}_a(\mbox{con})\equiv \vec{E}(\mbox{inside})$ is only present
in the surface at $x=0$ of the conductor due to the skin effect; 
$\vec{E}_a(\mbox{con})=-g_{a\gamma\gamma}a_0\vec{B}\exp(-x/\delta_e)\cos(m_at-x/\delta_e)$ with skin depth $\delta_e$.
It satisfies $\vec{E}_a(\mbox{con})(x=0)=\vec{E}_a=-g_{a\gamma\gamma}a(t)\vec{B}$.
( We assume that the flat conductor occupies the region $x>0$ and 
extends infinitely in $y$ and $z$ directions. Its surface is located at $x=0$. 
We assume that the magnetic field $\vec{B}$ is imposed parallel to the surface of the conductor. )
The electric field induces electric current $\vec{J}_a=\sigma \vec{E}_a(\mbox{con})$
in the surface. 
The current is enhanced in comparison with
the electric current $\vec{J}_v$ in vacuum induced by axion, $\vec{J}_v=g_{a\gamma\gamma}\partial_t a(t)\vec{B}
=-g_{a\gamma\gamma}m_aa_0\sin(m_at)\vec{B}$. ( The current can be derived from the term in the left hand side of 
the second equation in eq(\ref{metal}). )
That is, the strength of the current is given by 
$J_a=\sigma g_{a\gamma\gamma}a_0B = J_v/(m_a\delta_e)^2$ with the skin depth $\delta_e=\sqrt{2/(m_a\sigma)}$
and $J_v\equiv g_{a\gamma\gamma}m_aa_0B$.
In our previous papers we have calculated radiations
from the electric current $\vec{J}_a$ by using standard formula of electromagnetic gauge fields,

\begin{equation}
\label{1}
\vec{A}=\frac{1}{4\pi}\int \frac{\vec{J}_a(\vec{x}', t-|\vec{x}-\vec{x}'|)}{|\vec{x}-\vec{x}'|}d^3x' .
\end{equation}
which satisfy the equations $(\partial_t^2-\vec{\partial}^2)\vec{A}=\vec{J}_a$ describing electromagnetic fields $\vec{A}$ 
emitted by the current $\vec{J}_a$.
The time component $A_0$ is derived by solving
Landau gauge condition, $\partial\cdot\vec{A}+\partial_t A_0=0$. 
%

Because the electric current density $J_a =J_v/(m_a\delta_e)^2$ is much larger than 
the current density $J_v$, the radiation is expected to be enhanced. Actually, we can see from the formula of $\vec{A}$ in eq(\ref{1}) that
the field strength of the radiation turns out to be of the order of $E_a/(m_a\delta_e)\sim 10^4E_a$ because $1/(m_a\delta_e)\sim10^4$,
where $E_a\equiv g_{a\gamma\gamma}a_0B$.
That is, 
it is larger by $10^4$ times than the electric field $E_a$ in vacuum induced by axion.
It apparently seems that the idea mentioned above is reasonable.

\vspace{0.1cm}
We may ask whether or not there are something wrong in the above argument.
In the above argument, we have not taken into account the effect of radiation on the electric field $\vec{E}(\mbox{con})$.
The radiation has electric field $\vec{E}(\mbox{outside})$ which can affect the electric field through the boundary condition,
$\vec{E}(\mbox{outside})=\vec{E}(\mbox{con})$. 
Furthermore, 
we note that the electric current $\vec{J}_a=\sigma \vec{E}_a(\mbox{con})$ increases more as the conductivity $\sigma$ increases larger. 
So, the strength of the radiation increases. But we know that the electric field inside the conductor must vanish
in the limit of the perfect conductor $\sigma \to \infty $. It does not hold for $\vec{E}_a(\mbox{con})(x=0)\neq 0$.
We also know that total electric fields of the radiation $\vec{E}(\mbox{outside})$ and $\vec{E}_a$ induced by axion 
in vacuum also vanishes at the surface in the limit. 
( The radiation naively expected has an electric field component $\vec{E}(\mbox{outside})$ 
of the order of $10^4E_a$. It never cancels the electric field $\vec{E}_a$
in vacuum induced by axion. )

The electric field in vacuum parallel to the surface of the conductor
must be continuous at the surface. ( Electric field longitudinal to the surface also must be continuous because there are no electric charges
on the surface. ) Therefore, we naively expect that real electric field $\vec{E}(\mbox{inside})$ inside the conductor decreases as the conductivity 
becomes larger. As a result, the electric field $\vec{E}(\mbox{inside})$ vanishes for $\sigma \to \infty $. Then, both of the electric field present in vacuum and
$\vec{E}(\mbox{inside})$ satisfy the condition of their continuity at the surface. Hence,
we understand that the electric field $\vec{E}_a(\mbox{con})$ naively induced by axion is not real electric field $\vec{E}(\mbox{inside})$ present in the conductor.
It is suppressed in the conductor. The suppression is caused by radiation from the oscillating current in the surface.
According to the suppression of the electric field $\vec{E}_a(\mbox{con})$, real electric current $\vec{J}$ emitting the radiation is
not $\vec{J}_a=\sigma \vec{E}_a(\mbox{con})$. The electric current $\vec{J}_a$
in the formula of radiation $\vec{A}$ has to be replaced by the real current density $\sigma \vec{E}(\mbox{inside})$.

\vspace{0.1cm}
We will explain the suppression in the following. 
Suppose that a slab of conductor is imposed by magnetic field $\vec{B}$ parallel to its surface. 
The field $\vec{B}=(0,0,B)$ is pointed to $z$ direction and the surface of the slab is located at $x=0$. It occupies the region $x \ge 0$.
The slab with electrical conductivity $\sigma$ is homogeneously extended infinitely in $y$ and $z$ directions.
Then, the electric field $\vec{E}_a=-g_{a\gamma\gamma}a(t)\vec{B}\equiv (0,0,-E_a\cos(m_at))$ is induced by axion in vacuum $x<0$.
We can see from the equations (\ref{metal}) that
electric field inside the conductor $x \ge 0$ satisfies the equation, 
$(\vec{\partial}^2-\partial_t^2)\vec{E}(\mbox{inside})=\sigma\partial_t\vec{E}(\mbox{inside})-\partial_t^2\vec{E}_a$. 
The first term in right hand side of the equation represents the time derivative $\partial_t\vec{J}$ of the electric current 
$\vec{J}=\sigma \vec{E}(\mbox{inside})$ and
the last term represents the effect of axion.
The solution
is generally given such that

\begin{equation}
\label{2}
\vec{E}(\mbox{inside})= \vec{E}_1\exp(-\frac{x}{\delta_e})\Big(\cos(m_a t-\frac{x}{\delta_e})+A\sin((m_a t-\frac{x}{\delta_e})\Big)
+\frac{1}{\sigma}\partial_t \vec{E}_a,
\end{equation}
where we have explicitly written the terms involving $\sin(m_a t-\frac{x}{\delta_e})$ and $\cos(m_a t-\frac{x}{\delta_e})$ for convenience. 
This is because the electric field $\vec{E}_a$ in vacuum is defined to has the form of $-g_{a\gamma\gamma}a_0\cos(m_at)\vec{B}$. 
The cosine has no arbitrary phase.
The electric field $\vec{E}_a$ in vacuum 
determines the field $\vec{E}(\mbox{inside})$ inside the conductor. That is,
the coefficients $\vec{E}_1=(0,0,-E_1)$ and $A$ are determined by
the boundary conditions at the surface of the conductor. The first term in eq(\ref{2}) represents oscillating electric field 
only present in the surface with the skin depth, 
while the second term
is present even inside the conductor. 
This term $\frac{1}{\sigma}\partial_t \vec{E}_a$ is not present
in the slab with finite length in the $z$ direction owing to the screening by electric charges induced on
two ends of upper and lower surfaces. So we neglect it hereafter.

\vspace{0.1cm}
When there is no radiation from the slab, only electric field $\vec{E}_a$ is present in vacuum. Then, the coefficients are determined such that
$E_1=E_a$ and $A=0$ owing to the boundary condition, $\vec{E_a}=\vec{E}(\mbox{inside})$ at $x=0$.
Obviously, the electric field $\vec{E}(\mbox{inside})=\vec{E}_a(\mbox{con})$ generates oscillating electric current $\vec{J}_a=\sigma \vec{E}_a(\mbox{con})$
in the surface of the conductor. It is naive one mentioned above.
The strength of the current $J_a$
is enhanced, compared with the vacuum current $J_v$; $J_a\sim J_v/(m_a\delta_e)^2$
with $1/(m_a\delta_e)^2 \sim 10^8$.
Thus, the enhanced radiation is expected to arise. 

The estimation does not involve the
effect of radiation on the electric field inside the conductor. 
The effect causes the suppression of the electric field $\vec{E}_a(\mbox{con})$.
Actually, 
in the presence of the radiation, 
we must take into account the electric field of the radiation in addition to the electric field $\vec{E}_a$
in the boundary condition.
We recalculate the electric field $\vec{E}(\mbox{inside})$ inside the conductor by
taking account of the radiations from the conductor.
Because the radiation is outgoing wave, it is described as the wave of electric and magnetic fields,

\begin{eqnarray}
E_z(\mbox{outside})&=&b_1\cos m_a(t+x)+b_2\sin m_a(t+x) \\
B_y(\mbox{outside})&=&b_1\cos m_a(t+x)+b_2\sin m_a(t+x)
\end{eqnarray}
where we have written down only non vanishing components. Both of $E_z(\mbox{outside})$ and $B_y(\mbox{outside})$ are identical to each other
because of a Maxwell equation $\vec{\partial}\times \vec{E}+\partial_t \vec{B}=0$, which leads to
the equation $\partial_xE_z(\mbox{outside})=\partial_t B_y(\mbox{outside})$. 

In addition to the electric field $\vec{E}(\mbox{inside})$, 
we have magnetic field inside the conductor corresponding to the electric field $\vec{E}(\mbox{inside})$
( $\partial_xE_z(\mbox{outside})=\partial_t B_y(\mbox{outside})$ ),

\begin{eqnarray}
\label{mag}
B_y(\mbox{inside})&=&\frac{E_1}{\delta_e m_a}\exp\big(-\frac{x}{\delta_e}\big)\Big(\sin(m_a t-\frac{x}{\delta_e})
+\cos(m_a t-\frac{x}{\delta_e})\Big) \nonumber \\
&+&\frac{AE_1}{\delta_e m_a}\exp\big(-\frac{x}{\delta_e}\big)\Big(\sin(m_a t-\frac{x}{\delta_e})
-\cos(m_a t-\frac{x}{\delta_e})\Big).
\end{eqnarray}
  
Therefore, by imposing the boundary conditions on electric field $\vec{E}_a+\vec{E}_z(\mbox{outside})=\vec{E}(\mbox{inside})$  
and magnetic field $\vec{B}(\mbox{outside})=\vec{B}(\mbox{inside})$ at $x=0$,
we find  

\begin{equation}
E_a-\frac{E_1-AE_1}{m_a\delta_e}=E_1 \quad  A E_1=-\frac{E_1+AE_1}{m_a\delta_e} \quad \mbox{and} \quad
b_1=-E_1+E_a \quad b_2=-A E_1.
\end{equation}
By noting $m_a\delta_e \ll 1$, we obtain the solutions,
$E_1\simeq m_a\delta_e E_a/2,\,\,  A\simeq -1+O(m_a\delta_e),
\,\, b_1\simeq E_a(1+O(m_a\delta_e))$ and $b_2=m_a\delta_e E_a/2$.
That is,

\begin{eqnarray}
\vec{E}(\mbox{inside})&=&(0,0, -\frac{m_a\delta_e E_a}{2})\exp(-\frac{x}{\delta_e})\Big((\cos(m_at-\frac{x}{\delta_e})-\sin(m_at-\frac{x}{\delta_e})\Big)\\
E_z(\mbox{outside})&\simeq &E_a\cos m_a(t+x) \\
B_y(\mbox{outside})&\simeq &E_a\cos m_a(t+x)
\end{eqnarray}

Therefore, we find that the strength of outgoing electromagnetic field emitted from the conductor is of the order of $E_a$
and that the electric current $\vec{J}=\sigma \vec{E}(\mbox{inside})\simeq (0,0,-m_a\delta_e \sigma E_a/2)$ inside the conductor is much less than
$\sigma E_a$ estimated naively above.

Furthermore, the outgoing radiation satisfies $\vec{E}_a+\vec{E}(\mbox{outside})=\big(0,0,-E_a+E_z(\mbox{outside})\big)=0$ at $x=0$ in the limit of
perfect conductor, $\delta_e\to 0$ ( $\sigma \to \infty$ ).
Similarly we have $\vec{E}(\mbox{inside})=0$ in the limit.
Therefore, we find that the electromagnetic field emitted by the conductor has the same order of magnitude as the electric field 
$\vec{E}_a$ induced by axion in vacuum. 
Although the radiation from the flat conductor is never enhanced, the time averaged total flux of the radiation can be large
such that $\overline{E_zB_y}\times S=E_a^2S/2\sim10^{-26}\rm W(S/100\rm m^2)(\rho_a/0.3\rm GeVcm^{-3})(B/1\rm T)^2$ ( $S$ surface area of the 
conductor ).

\vspace{0.1cm}
Physically, we interpret the fact in the following. The dark matter axion under homogeneous magnetic field produces
electric field $\vec{E}_a=-g_{a\gamma\gamma}a_0\cos(\omega_a t +\vec{k}\cdot\vec{x})$ with small momentum $k\sim 10^{-3}m_a$
with $\omega_a=\sqrt{m_a^2+k^2}\simeq m_a+k^2/(2m_a)$. 
( In the above discussion we have neglected the small momentum $k$ of the order of $10^{-3}m_a$. )
It propagates and collides the slab. The wave is reflected by the slab and the reflected wave is
the outgoing one
described by $E_z(\mbox{outside})$ and $B_y(\mbox{outside})$ shown in the above calculation.
It is reasonable that the amplitude of the outgoing wave is identical to that of the incoming wave $E_a$.
It is the physical reason why the outgoing radiation from the conductor is not enhanced.

\vspace{0.2cm}
The author
expresses thanks Izumi Tsutsui and Osamu Morimatsu for serious comments and discussions.
He also expresses special thank to Yasuhiro Kishimoto and Kazunori Nakayama for 
indicating author's misunderstanding in the previous papers. They have obtained the similar result to ours.

This work is supported in part by Grant-in-Aid for Scientific Research ( KAKENHI ), No.19K03832.

\end{document}